**Strong, Tough, and Stiff Bioinspired Ceramics from Brittle Constituents**

Florian Bouville [1,2], Eric Maire [2], Sylvain Meille[2], Bertrand Van de Moortèle [3], Adam J. Stevenson [1], Sylvain Deville [1]

[1] Laboratoire de Synthèse et Fonctionnalisation des Céramiques, UMR3080 CNRS/Saint-Gobain, Cavaillon, France

[2] Université de Lyon, INSA-Lyon, MATEIS CNRS UMR5510, Villeurbanne, France

[3] Laboratoire de Géologie de Lyon, Ecole Normale Supérieure de Lyon, Lyon, France

**High strength and high toughness are usually mutually exclusive in engineering materials. Improving the toughness of strong but brittle materials like ceramics thus relies on the introduction of a metallic or polymeric ductile phase to dissipate energy, which conversely decreases the strength, stiffness, and the ability to operate at high temperature. In many natural materials, toughness is achieved through a combination of multiple mechanisms operating at different length scales but such structures have been extremely difficult to replicate. Building upon such biological structures, we demonstrate a simple approach that yields bulk ceramics characterized by a unique combination of high strength (470 MPa), high toughness (22 MPa.m$^{1/2}$), and high stiffness (290 GPa) without the assistance of a ductile phase. Because only mineral constituents were used, this material retains its mechanical properties at high temperature (600°C). The bioinspired, material-independent design presented here is a specific but relevant example of a strong, tough, and stiff material, in great need for structural, transportations, and energy-related applications.**

Ceramics exhibit among the highest stiffness and strength of all known material classes[1]. Because of the strong and directional bonding between constitutive atoms, they present a high fusion temperature and thus a high thermal stability. Being composed of mostly light elements, they are also lightweight compared to most metals. This property combination predestined them to be the best material choice for high stress and high temperature operative conditions. However, the iono-covalent bonds limit most common plasticity mechanisms and therefore prevent any ductile behavior. This lack of plasticity is often the main limitation for the use of ceramic materials, resulting frequently in catastrophic and unpredictable failure[2] which greatly limits their range of applications. Damage-resistant ceramics are thus in great demand.

However, countless damage-resistant biological materials comprise ceramics and are used for structural purposes[3,4] in nature. The most famous example is probably the nacreous part of seashells where the brick and mortar structure comprising 95 vol.% of platelets of polycrystalline aragonite ($CaCO_3$) and 5 vol.% of a protein is extremely effective at restricting crack propagation[5] and presents a toughness orders of magnitude greater than either of its constituents[6,7]. Nacre toughness is achieved through numerous extrinsic mechanisms that include viscoelastic deformation of the protein layer[8], mineral bridges rupture, and inelastic shearing and frictional sliding during platelets pull out[5].

Tough ceramic materials could be obtained by translating the specific microstructural and interfacial structures present in natural systems to engineering materials[9]. This could be achieved by building "brick and mortar" organized structures using anisotropic particles as building blocks combined with a ductile phase. The alignment of anisotropic particles can be achieved by sequential deposition methods[10] for thin films but not for bulk materials.

Ceramic/polymer composites that replicate nacre features have been fabricated by various techniques, such as ice templating, magnetic particles alignment[9], or simply by gluing ceramic layers[11] and the toughness levels achieved are rather impressive[12]. Process using magnetic fields lead to structures with well-aligned platelets embedded in a polyurethane matrix[13]. Such approaches nevertheless suffer from severe limitations. Because the design is based on the presence of a ductile phase, the materials exhibiting the highest toughness all comprise at least 20 vol.% of a polymeric phase, restricting *de facto* their useful temperature range to mild temperatures (200°C or less). In nacre and nacre-inspired materials, the intrinsic toughening is primarily a result of crack deflection at the platelets interfaces. Thus, as a first order approximation, increasing the local density of interfaces should increase the toughness of the material. Such increase can be achieved by the use of small constituents. Finally, the processing route still includes too many steps to make it viable beyond the lab scale.

Here we demonstrate the use of a relatively simple processing route based on widespread ceramic processing techniques to obtain fully ceramic, layered bulk materials with submicrometer layer spacing that exhibit not only an unprecedented toughness for a ceramic material but also, because it comprises only mineral constituents, a very high strength and stiffness even at high temperature.

**Processing strategy**

Inspired by structural guidelines derived from the study of nacre, we fabricated dense ceramic materials defined by five structural features, spanning several length scales (Fig. 1A): (i) long range structural order, (ii) closely packed ceramic platelets of dimensions identical to that of nacre, (iii) ceramic bonds (bridges) linking the platelets, (iv) nano-asperities at the surface of platelets, and (v) a secondary phase with lower stiffness ensuring load redistribution, crack deflection, and delamination.

The process we used here is based on ice-templating. However, we did not rely on ice crystals to form the elementary bricks, but instead took advantage of their growth as a driving force for the local self-assembly of anisotropic particles (platelets). Those particles have the same dimensions as nacre platelets (500 nm thickness, 7 μm diameter) and constitute the elementary building blocks of the structure. When a suspension is directionally frozen, the metastable growth of the ice crystals[14] repels and concentrates the particles present in the suspension. The concentration of the particles occurs at a length-scale where self-assembly of platelets can occur[15] (Fig. 1A). Alumina nanoparticles (100 nm) incorporated in the initial suspension serve as a source of both inorganic bridges between the platelets and nanoasperities at the surface of platelets, similar to that observed in nacre. Finally, smaller nanoparticles (20 nm) of liquid phase precursors (silica and calcia) are added to aid filling the remaining gaps during the sintering stage. The material is thus composed of 98.5 vol.% alumina, 1.3 vol.% of silica, and 0.2 vol.% of calcia. This simple strategy, where all the constituents are incorporated in the initial suspension and self-organized in one step, allows precise and easy tailoring of the final material composition. Long-range order of the ice crystals is obtained through a freezing under flow method (see Methods). Once the porous samples are obtained, they are simply pressed and sintered by field assisted sintering (FAST). The materials are 86% porous after the freeze drying step. The pressing step (100MPa) priori to sintering is critical to remove almost all of the macroporosity. This is easily achieved thanks to the long-range order of the macropores induced by the controlled freezing step.

Due to their size and high aspect ratio, the platelets themselves are unlikely to densify by sintering. Instead, the liquid phase formation rearrange the particles through capillary forces, facilitating platelet packing under the applied load by lubricating the contact points, and by filling the pore space between them (Fig. 1B). At the same time, the nanometric alumina particles form strong bridges by sintering to the platelets. The high surface area and surface curvature of nanoparticles allow them to sinter at lower temperatures than the platelets[16] (Fig. 1C). One important consideration is ensuring that the alumina nanoparticles are not completely dissolved in the liquid phase by either limiting the total amount of liquid phase, or by using short processing times like those employed here (enabled by the rapid, pressure assisted densification method). The sample with liquid phase starts shrinking at a higher temperature (Fig. 1C) than the sample without the liquid phase. We selected this glass composition (calcia and silica) for its very high melting temperature (pure silica start melting at around 1600°C). We believe that the precursor particles of the glass phase prevent most of the contacts between alumina platelets and the nanoparticles, and thus delay the densification of the alumina nanoparticles. The glass has to melt first before densification can proceed. In addition, because the liquid phase precursors are well dispersed in the system and in low concentration compared to the alumina platelets and nanoparticles, the local composition throughout the green body is not likely to be the liquid phase composition predicted from the phase diagram - there must be some diffusion and densification before the liquid phase formation. Along with the physical restraint described above, this may contribute to the slower densification kinetics observed.

The entire process thus consists of ~~just~~ three steps: (i) preparation of an aqueous colloidal suspension, containing all the required building blocks and processing additives, (ii) ice templating this suspension, and (iii) pressure assisted sintering step at 1500°C. The materials obtained are referred to as nacre-like alumina.

**Nacre vs. nacre-like alumina microstructure**

The final microstructure of the nacre-like alumina (Fig. 2) is characterized by a dense packing of platelets presenting long range order. The use of a fast sintering method limits grain growth so that the initial dimensions of the platelets are retained. A bulk material with submicronic layer spacing is thus obtained, a feature of nacre that had proved impossible to replicate so far in bulk materials. Compared to the microstructure of nacre (Fig. 2, A and B), the packing of platelets presents short range order (Fig. 2E), but the long range order is not as perfect. The waviness of the stacking (Fig. 2D) in our nacre-like ceramic, estimated at ± 15° around the main orientation, comes from the organization of the ice crystals. The long range order is nevertheless sufficient to obtain nearly fully dense samples (e.g., >98% of relative density). The relative density calculated by Archimedean method are very similar: platelets and nanoparticles: 98.9%, platelets and liquid phase: 98.4%, and platelets, nanoparticles and liquid phase: 98.8%.

Observations of the interfaces between platelets (Fig. 2, E and F) reveal the presence of a secondary phase mimicking the organic layer in nacre (Fig. 2B). It also reveals some alumina bridges between adjacent platelets and finally nanoasperities analogous to those in nacre (Fig. 2C). Based on the experimental conditions and the ceramic phases in presence, this secondary phase is most likely amorphous. A small fraction of the alumina platelets and nanoparticles is dissolved into this glass phase at the interface, so that the actual final fraction of this phase (estimated to be around 2 to 3 vol.%) is slightly higher than initially incorporated into the suspension (1.5 vol.%). We estimated the fraction of the glass phase based on the amount of glass precursors (initially 1.5 vol.%) and nano particles incorporated (initially 3 vol.%), and the corresponding ternary phase diagram (alumina-silica-calcia). Based on this phase diagram[17], we can estimate the composition of the glass phase at 10 wt.% $SiO_2$, 65 wt.% CaO, and 25 wt.% $Al_2O_3$, although it is hard to tell whether equilibrium is achieved during the rapid densification.

The microstructure of the synthetic material is thus strikingly similar to that of natural nacre at several length scales, validating the possibility to fabricate bulk, centimeter-sized samples with only a few simple processing steps.

**Ductile ceramics?**

Single-edge notched beam (SENB) tests were carried out to measure the toughness and the behavior of the materials tested was compared with a reference polycrystalline alumina sample. The resulting stress-strain curves are plotted in Fig. 3A for three model compositions to illustrate the synergistic effects between the structural features introduced: platelets with nano-asperities, platelets with a liquid phase, and platelets with both (the nacre-like alumina structure). As expected, the samples containing the liquid phase only at the platelet interface exhibit a purely linear elastic response until a catastrophic failure occurs, with unstable crack propagation characteristic of brittle materials. The corresponding fracture surface, showing no crack deflection at all, is shown in Fig. 3D. The fracture toughness $K_{Ic}$ (6.1 MPa.m$^{1/2}$) is nevertheless already significantly higher than a reference alumina (3.5 MPa.m$^{1/2}$) since platelets are significantly harder to separate than isotropic particles due to their orientations[18]. Platelets with nano-asperities exhibit a very significant toughening stage. Progressive failure occurs by stable crack propagation, a very unusual behavior in pure ceramic materials.

When both reinforcements are combined, leading to the nacre-like alumina structure, we obtain stable crack growth combined with toughening, similar to that of nacre (reported for comparison in Fig. 3B). The nacre-like alumina, free of any ductile component, is clearly stiffer, with flexural modulus of 290 GPa (compared to 40 GPa for nacre). The mechanical response, where more than 50% of the strain to failure is inelastic, is strikingly reminiscent of that observed in ductile metallic or organic materials, a remarkable and unexpected behavior for a material exclusively composed of ceramic constituents. The reinforcement mechanisms (discussed in the next section) are extrinsic, which means that of course no true ductility (plastic deformation without crack propagation) is observed.

The fracture toughness $K_{Ic}$ describes only the resistance to a crack initiation and is significantly higher in this nacre-like structure (6.2 MPa.m$^{1/2}$) than in usual polycrystalline alumina (3.5 MPa.m$^{1/2}$). It is nevertheless unable to describe the multiple extrinsic toughening mechanisms that take place during crack propagation and greatly contributes to the dissipation of energy. The crack is deflected by the composite low stiffness interface[19] (Fig. 3E) and slowed down by various extrinsic toughening mechanisms which result in stable crack growth, similar to natural nacre[5]. This toughening leads to an increase of the fracture resistance as the crack propagates. This behavior, known as an R-curve effect, is characteristic of many natural materials such as nacre[5], bone[3,4], and tooth[20]. In order to measure the R-curve, the indirect crack length is measured by a compliance method (see Methods). Crack extension reported here satisfied a small-scale bridging conditions (see Methods). Analogous approaches have been used previously in similar structures[2,3,12] where multiple extrinsic toughening mechanisms needed to be described. The R-curves at room temperature and high temperature (600°C) are plotted in figure 3C. By taking into account the local deflection as well as the other dissipation mechanisms with J-integral and using the equivalence in stress intensity factor, the maximum increase of toughness is extremely high, around 22 MPa.m$^{1/2}$. This corresponds to a 350% increase compared to the $K_{Ic}$ toughness (600% increase with respect to the reference alumina). This far exceeds that of nacre and is equivalent to the best brick and mortar polymer/ceramic composites developed previously[12]

The second stage of crack growth, when the crack length exceeds 400 µm, is unstable. Nevertheless, different energy dissipating mechanisms are present, eventually leading to the arrest of cracks before catastrophic failure can occur. Notch-free samples are not fully broken after being exposed to their maximum stresses, another evidence of the efficiency of the reinforcement in these materials.

The properties are retained at relatively high temperature (600°C, Fig. 3B), with a stress intensity factor for crack initiation ($K_{Ic}$) of 4.7 MPa.m$^{1/2}$ and a $K_{JC}$ maximum toughness of 21 MPa.m$^{1/2}$ (Fig. 3J).

**Toughening mechanisms**

We combine with these materials a number of toughening mechanisms, operating at different length scales. By restricting our choice of constituents to ceramic materials to ensure their use at high temperature and maximize the stiffness, we are also eliminating most of the usual toughening mechanisms seen in ductile materials such as dislocation plasticity or shear bands. The extrinsic reinforcement developed here is derived from the hierarchical architecture of the material and the load redistribution resulting from the brick and mortar architecture. Crack deflection (Fig. 3E), multiple cracking, crack bridging, and crack branching (Fig. 3F), and delamination (Fig. 3H) operate at the crack tip, effectively relieving the locally high stresses. The alumina nanoparticles provide both bridges between the platelets and nano-asperities. This leads to efficient energy dissipation by frictional sliding during platelets pull-out and breaking of the bridges. Pores located between adjacent platelets, as seen in Fig. 2F, may also contribute by acting as local crack arresters by a Cook-Gordon mechanism[21]. The possibility to achieve toughness through a proper combination of brittle constituents has been recently hypothesized[22]. With the proper topological arrangement and stiffness ratio between two brittle phases, the stress concentration around the crack tip can be reduced thanks to more efficient load redistribution. The deformation is delocalized and strain redistributes throughout the entire structure. These mechanisms are likely to contribute to the level of toughness achieved in these structures. Further work will be necessary to quantify the relative contributions of all the toughening mechanisms identified here.

**A high strength, high toughness material**

The flexural strengths of our materials are reported in Fig. 3I. Platelets combined with either the liquid phase or the nano-asperities alone have, as expected, a flexural strength equivalent to or lower than the reference alumina. The nacre-like alumina exhibits a flexural strength of 470 MPa at room temperature and 420 MPa at 600°C, a value essentially similar to that of the reference alumina. The presence of pores (defects) in the platelets and nanoparticles material may explain its lower strength. The synergistic combination of both the liquid phase and the nano-asperities is thus strikingly demonstrated by these measurements. The strength is similar in the transverse direction (490 MPa perpendicular to the basal plane of the platelets). This material does not present strength anisotropy and thus any weakness in these two solicitation directions. Because of sample size limitations (maximum thickness around 5mm), the strength in the delamination direction cannot be assessed.

Strength and toughness are generally considered to be mutually exclusive[2]. Because intrinsic toughening mechanisms are linked to plasticity and thus strength, a compromise is always reached in structural materials where either one of the properties is set aside. In similar work[12] for instance, the introduction of 20 vol.% of a polymeric phase to increase the toughness resulted in a notable decrease of the flexural strength. Further toughness increases can be achieved through the addition of up to 60 vol.% Al-Si alloy[23], but no appreciable gains in strength are obtained. The introduction of a polymer or a metallic phase is thus always done at the expense of strength, stiffness, and temperature range of use. The toughness of the nacre-like alumina is greater than other reported values for alumina[24], textured alumina[25], or even for alumina composites[26], composed of either an organic phase[12], a metallic phase[27–31], or carbon reinforcements such as carbon nanotubes[32–34] or graphene[35] (Fig. 4A). As in nacre, the organization of the microstructure induces the emergence of improved properties without compromise, even at elevated temperature.

When it comes to materials choice, in particular for structural applications, the density is a major consideration. Metallic glasses are some of the toughest engineered materials produced so far but their high density limits their use. Thickness is also a major problem because the necessary cooling rates cannot always be achieved on bulk samples. The unique combination of specific strength ($\sigma_f/\rho$) and specific toughness ($K_c/\rho$) of our bioinspired ceramic material actually matches that of engineering aluminum and magnesium alloys (Fig. 4B) while exhibiting a higher hardness (16 GPa), stiffness, and operating temperature. Because only ceramic constituents are used, the stiffness, strength and operating temperature of our material are at the usual level of engineering, technical ceramics, highlighting their remarkable potential in structural and energy-related applications.

For now, the dimensions of the samples are limited to a few centimetres in diameter and a few millimetres in thickness. Two steps of the process induce size constraints: freezing and sintering. Although ice-templating has been restricted to the laboratory scale so far, there are no fundamental reasons why the process could not be scaled up. Large scale, low temperature processes are widespread in the industry (frozen food being the most notable example). The sintering process is actually very similar to Hot Pressing, except that an electric field is used to heat rapidly the graphite die. This kind of pressure-assisted equipment, assisted by an electric field, is now routinely used and widespread, pieces up to 80 cm diameter are routinely sintered for a wide range of applications. With a few exceptions such as refractory materials for furnaces, which can have a thickness of half a meter or so, technical ceramics pieces are almost always within this range of dimensions or even smaller. The majority of technical ceramics pieces have dimensions in the millimetre range. We therefore believe that this approach should fulfil the requirements for most structural applications of ceramics, such as molten metal processing and forming or high performances engines.

On a longer term, we believe that the design principles shown here are not restricted to ice templating, but could be adapted to standard colloidal processing routes for specific shapes, sizes, or directionality. We are currently investigating such ideas. The main issue we can expect, if standard colloidal processing routes are to be used, is how to control the orientation of anisotropic particles.

**Concluding remarks**

The bioinspired, material-independent design presented here is a specific but relevant example of a strong, tough, and stiff material, in great need for structural, transportations, and energy-related applications where catastrophic failure is not an option. The flexibility in the choice of materials permits an independent assessment (and thus tailoring) of the role of each structural feature on the mechanical properties, which is not possible in natural materials. Beyond structural materials, we expect to see many other benefits of using ice crystals as a driving force for the bulk self-organization of elementary building blocks. We are now focused on extending the concepts demonstrated here to other materials combinations.

**Methods summary**

**Suspensions preparation.** A system composed of alumina platelets (7 μm diameter, 500 nm thickness), alumina nanoparticles (diameter around 100 nm), and a silica-calcia liquid phase (diameter around 20 nm) have been used. All the constituents where added in distilled water and then ball-milled for 24 hours, except the alumina platelets that have been added 3 hours before the end of the cycle to avoid any excessive breakage by the milling media.

**Freezing under flow.** To obtain dense structures by pressing the porous sample obtained after ice templating, further control of the crystal growth is needed to obtain parallel crystals over large dimension. We use a method in which we let the slurry flow on the cooling plate where freezing occurs, leading to a second perpendicular temperature gradient during the freezing. A 1°C/min cooling rate is used for all suspensions.

**Field Assisted Sintering (FAST).** The equipment used for sintering is a HPD25 device from FCT System GmbH. The linear shrinkage rates were obtained from the movement of the die during the sintering. A constant pressure of 100 MPa was applied during the heat treatment. An electrical field is applied and mainly used to heat the graphite die by Joule effect. Due to the intensity of the input current, this method provides a high heating rate (around 100°C/min). Whether or not the current goes trough the sample is not completely clarified yet.

**Microstructural characterization.** SEM pictures were taken on uncoated samples by a Supra 55 microscope (Zeiss) and a ZEISS NVision40.

**Preparation of SENB and bending samples.** Beam shaped specimens where cut from the sintered disks and mirror polished. SENB test specimen were first notched with a diamond saw of 200 μm thickness and then the bottom of each notch was sharpened by repeatedly passing a razor blade with diamond paste.

**Hardness Measurement**: We used a MicroMet 6030 microdurometer to measure the hardness on mirror polished samples. The load applied was 0.3 kgf.

**Determination of crack length.** We used an equivalence between compliance and crack length on an SENB test. The crack length measured is thus the projection of the real crack length at the position of the notch. Samples deflection was measured by a linear variable differential transformer (LVDT) with micron accuracy. According to the ASTM criterion[36], the maximum crack extension is given by $\Delta a_{max}$=0.25b (b being the uncracked ligament width), which corresponds in our results to $\Delta a$= 0.4 mm. However, toughness measurements can be considered valid until the data becomes geometry dependent due to large scale bridging[37], which here corresponds to $\Delta a$= 0.8 mm. The values reported here were thus always obtained within a valid range of crack extension.

**Details on J-integral calculation.** To assess the different mechanisms that occurred during the stable crack propagation, a J-integral versus crack extension has been calculated as the sum of elastic and plastic contribution, a method already used to measure the properties of bone[3,38] and similar structures[12,37].

**Full Methods** are available in the online version of the paper.

**Supplementary Information** is available in the online version of the paper.

**Acknowledgements**: We acknowledge the financial support of the ANRT (Association Nationale Recherche Technologie) and Saint-Gobain through a CIFRE fellowship, convention #808/2010. We are indebted to the Centre Lyonnais de Microscopie (CLYM) for access to the FIB microscope. Acknowledgements are due to Guillaume Bonnefont from MATEIS for his assistance on the sintering equipment, and Catherine Barentin from the ILM for tipping us on the Carbopol to obtain a yield stress suspension.

**Authors Contributions:** SD and EM designed the research, FB processed the sample and performed the mechanical testing, FB and SM performed the high temperature mechanical testing, FB and BVdM investigated the structure, all authors analyzed and discussed the results, FB, AS, and SD wrote the paper.

**Author Information:** Reprints and permissions information is available at www.nature.com/reprints. The authors declare no competing financial interests. Correspondence and requests for materials should be addressed to S.D. (sylvain.deville@saint-gobain.com).

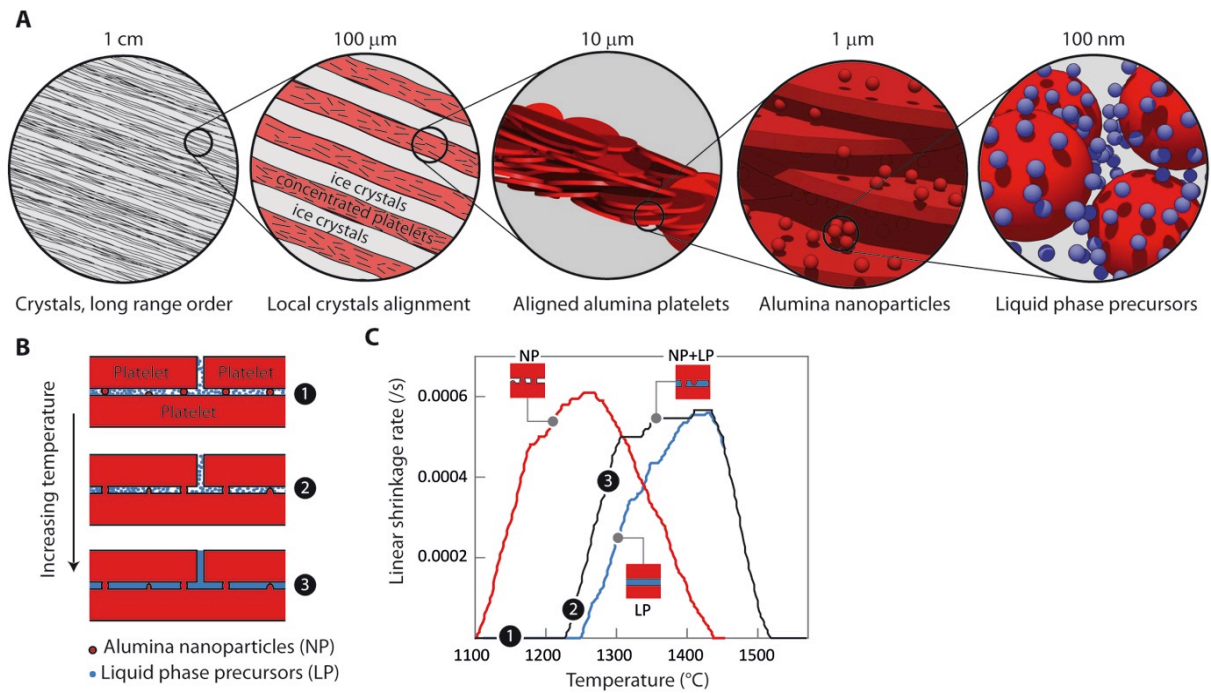

**Fig. 1. Design strategy describing the control at multiple scale of structural self-organization and the densification strategy. (A)** Self-organization of all the structural features occurs during the freezing stage. The growth of ordered-ice crystals triggers the local alignment of platelets. Alumina nanoparticles and liquid phase precursors are entrapped between the platelets. **(B)** Schematic representation of the densification scenario. **(C)** The linear shrinkage rate illustrates that the densification of the composition comprising either one of the building blocks (3 vol.% nanoparticles or 5 vol.% liquid phase) occurs at different temperatures. In the composition comprising all the building blocks, the densification starts at a temperature between the two others composition (liquid phase only and nanoparticles only) showing an interaction between nanoparticles and liquid phase precursors.

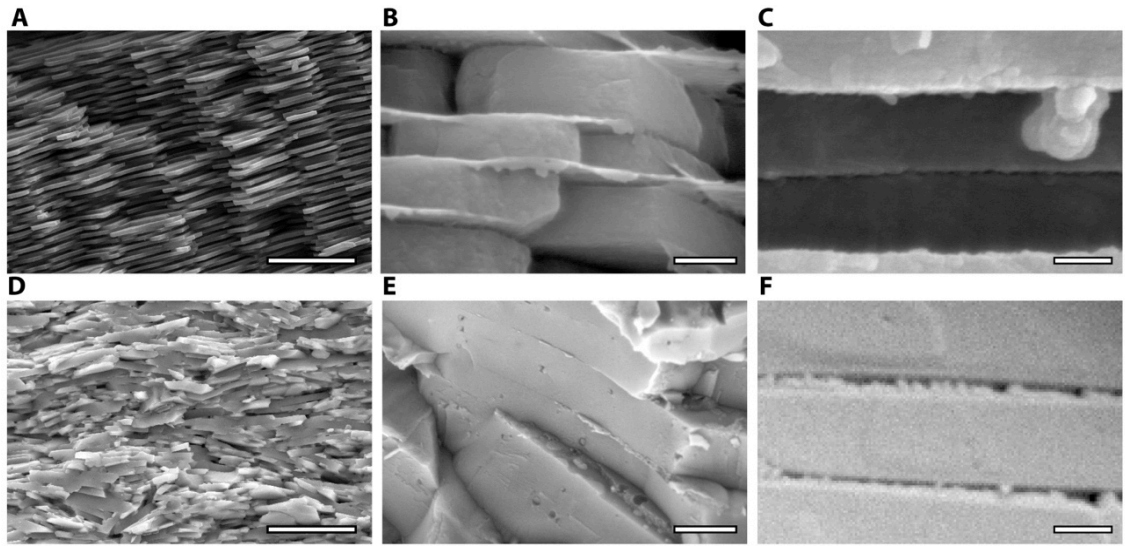

**Fig. 2. Comparison of the microstructure of nacre (A-C) and nacre-like ceramic (D-F).** (**A** and **D**) SEM micrograph showing the short and long range order of platelets. The nacre-like alumina shows relatively high organization of the platelets. (**B** and **E**), Local stacking of platelets. A liquid phase film is present even when the platelets are close, mimicking the protein layer in the nacre structure. (**C** and **F**) Closer views of platelets interface unveiling the presence of the inorganic bridges and nano-asperities along with the glassy (**F**) or organic (**C**) phase filling the space between adjacent platelets. Some residual pores are also visible. Scale bars: (**A**, **D**): 10 μm, (**B**, **E**): 500 nm, (**C**, **F**): 250 nm.

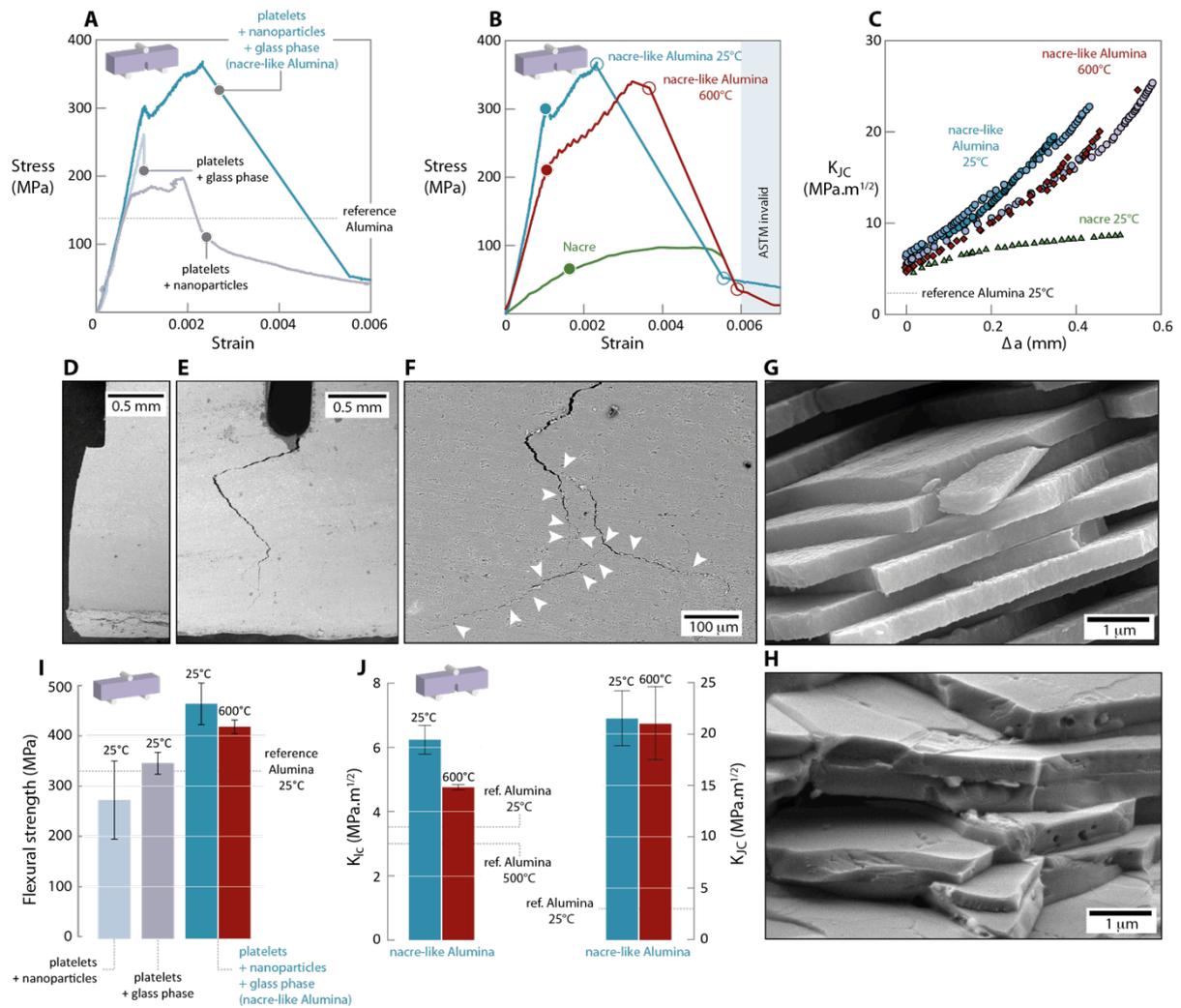

**Fig. 3. Mechanical properties and toughness measurement of nacre-like ceramic and nacre. (A)** Stress strain curve of the same three compositions on SENB samples. The 5 vol.% liquid phase presents brittle behavior, the 3 vol.% nanoparticles presents toughening and the combination leads to a very important toughening. **(B)** Stress strain curve for SENB samples of the nacre-like ceramic and natural nacre. The dots on the curve indicate the initiation of stable crack propagation. The circles indicate the beginning and the end of unstable crack propagation. **(C)** Fracture toughness $K_J$ calculated from J-integral and crack extension $\Delta a$ and the equivalent for the nacre-like ceramic and nacre. **(D)** Smooth fracture surface of brittle composition (platelets and liquid phase). **(E)** Fracture surface of a nacre-like sample that shows the long range crack deflection. **(F)** Multiple cracking and crack bridging towards the end of the crack path. Arrows indicate the onset of crack branching and bridging (**G** and **H**) Detail of the fracture surface in nacre (**G**) and nacre-like alumina (**H**) showing the crack deflection and delamination at the platelets interface. Scale bar: 1 μm. **I,** Flexural strength of three different compositions, liquid phase (5 vol.%) an platelets, nanoparticles (3 vol.%) and platelets, and nacre like alumina (respectively 1.5 vol.% and 3 vol.% of liquid phase and nanoparticles). The dot line corresponds to an equiaxed fine grain alumina on the **A** and **I** plots. Error bars indicate standard deviation. (J) Comparison of toughness for crack initiation ($K_{Ic}$) and stable crack propagation ($K_{JC}$) in nacre–like alumina and reference alumina at room temperature (25°C) and high temperature[39] (600°C).

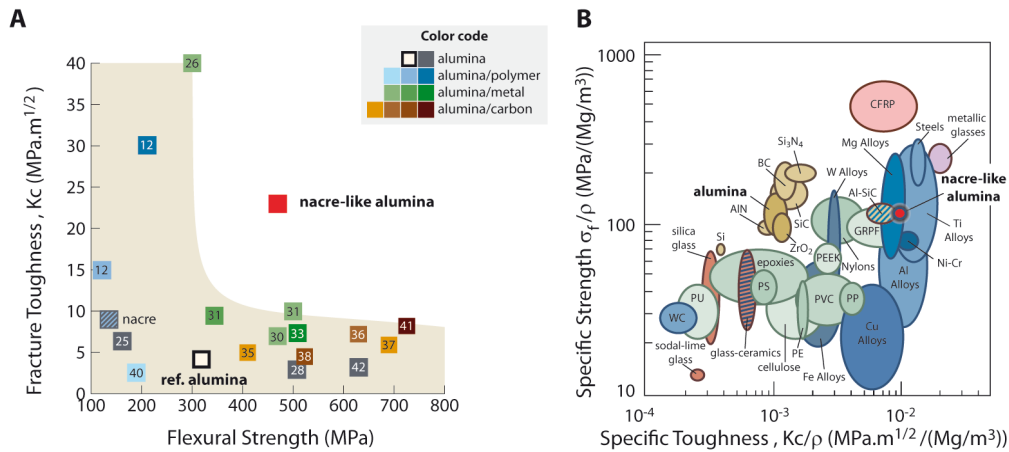

**Fig. 4: Comparison of the relative materials performance. (A)** The conflict between strength and toughness: fracture toughness vs. flexural strength, for alumina based materials. Improvements in toughness are obtained by the introduction of a ductile phase, such as a metallic phase[23,27–31], a polymer[12,40], or carbon reinforcements such as graphene[35], carbon nanotubes[32–34], or whiskers[41]. Improvements in strength are achieved through a texturation of the grains[25], hot isostatic pressing[42] or the addition of strong reinforcements. **(B)** Ashby diagram of specific strength versus specific toughness for a range of engineering and natural materials[43]. The nacre-like alumina's have specific strength/toughness properties similar to that of titanium or magnesium metallic alloys.

**Methods**

**Suspensions preparation**

All the constituents where added in distilled water and then ball-milled for 24 hours, except the alumina platelets that have been added 3 hours before the end of the cycle to avoid any excessive breakage by the milling media. A rheology modifier (Carbopol ETD 2691) was added to form a soft gel (yield stress suspension) in order to avoid any settling of the large particles during the freezing and thus ensuring the homogeneity of the sample throughout the specimen.

**Freezing under flow**

To obtain dense structures by pressing the porous sample obtained after ice templating, further control of the crystal growth is needed to obtain parallel crystals over large dimension. Indeed, the crystals are vertically aligned in the direction of the temperature gradient, but because nucleation occurs at different spots, different domains of lateral orientation are present[44]. To induce long range order of the crystals, different strategies have been developed[45]. Here we use a method in which we let the slurry flow on the cooling plate where freezing occurs, leading to a second perpendicular temperature gradient during the freezing. The crystals thus guided in two directions are all parallel, and we repeatedly obtain sample of relatively large dimensions (a few centimeters) with aligned crystals. A 1°C/min cooling rate is used for all suspensions.

**Field Assisted Sintering (FAST)**

The equipment used for sintering is a HPD25 device from FCT System GmbH. The sample is compressed between two graphite punches inside a cylindrical graphite die (20 mm diameter). The heating of the sample is obtained by injecting electrical current through the punches. This leads to a very high heating rate and the presence of the electrical current accelerates the sintering. The FAST device is known to allow sintering ceramics and metals at lower temperature than conventional sintering resulting in finer microstructures due to the applied pressure. As described, a system composed of alumina platelets (7 μm diameter, 500 nm thickness), alumina nanoparticles (diameter around 100 nm), and a silica-calcia liquid phase (diameter around 20 nm) have been used. The molar composition of 75:25 ($SiO_2$:CaO) have been chosen because of its high wettability of alumina surface and a relatively high melting point[25,46,47], between 1300°C and 1500°C. The densification behavior of different compositions is described in figure 1B. The linear shrinkage rates were obtained from the movement of the die during the sintering. All the relative density of the sample prepared here are above 98% after sintering at 1500°C with 100°C/min heating rate and a constant applied pressure of 100 MPa.

**Microstructural characterization**

SEM pictures were taken on uncoated samples by a Supra 55 microscope (Zeiss) and a ZEISS NVision40. The pictures were taken at low acceleration voltage (typically 1 kV) to avoid any charging effects.

**Preparation of SENB and bending samples**

Disk shaped samples with 20 mm diameter were obtained after sintering. Beam shaped specimens with dimensions around 14x2x2 mm$^3$ where then cut from the sintered disks. The beams for bend testing where mirror polished and beveled to avoid any crack departure from the sides. SENB test specimen were first notched with a diamond saw of 200 μm thickness and then the bottom of each notch was sharpened by repeatedly passing a razor blade with diamond paste (1μm). Using this method, the final notch radiuses were always below 40 μm.

At least 5 specimens were tested for each composition and setup. Mechanical testing were carried out on an INSTRON machine. Values were determined by monotonically loading the specimens to failure at a constant displacement rate of 1 µm.s$^{-1}$. The beams deflections were measured by a linear variable differential transformer (LVDT).

**Determination of crack length**

The indirect method usually used to determine crack length is based on complaisance evolution during cyclic loading. However in our case this method has proven to be unusable here because the repeated cycles applied induce small crack propagation, even at low stress. We instead used a simple equivalence between complaisance and crack length on an SENB test. The complaisance were calculated thanks to the relation $C=u/f$, where $u$ and $f$ are the displacement and force at each point after departure of a crack, respectively. Then the crack length was recursively calculated with (1).

$$a_n = a_{n-1} + \frac{W - a_{n-1}}{2} \frac{C_n - C_{n-1}}{C_n} \qquad (1)$$

where $W$ is the thickness of the specimen, $a$ and $C$ respectively the crack length and complaisance calculated at the n and n-1 step.

**Details on J-integral calculation**

To assess the different mechanisms that occurred during the stable crack propagation (crack length< 400 µm), a J-integral versus crack extension has been calculated as the sum of elastic and plastic contribution, a method already used to measure the properties of bone[3,38] and similar structures[12,37]. The elastic contribution J$_{el}$ is based on linear-elastic fracture mechanics (J$_{el}$=K$_{IC}^2$/E'),

The plastic component J$_{pl}$ is calculated with the relation (2):

$$J_{pl} = \frac{1.9 A_{pl}}{Bb} \qquad (2)$$

where $A_{pl}$ represents the plastic area under the load-displacement curve, $B$ the specimen lateral dimension and b the uncracked ligament.

A geometric mean of the local stress intensity factor leads to an equivalent stress intensity factor (3).

$$K_{JC} = \sqrt{(J_{el} + J_{pl})E} \quad (3)$$

with $K_{JC}$ the back calculated stress intensity factor, $J_{el}$ and $J_{pl}$ elastic and plastic contribution of the J-integral and $E$ the young modulus of alumina.